\documentclass[12pt]{iopart}

\usepackage{graphicx}
\usepackage{dcolumn}
\usepackage{bm}
\usepackage{amssymb}
\begin{document}

\title[The Nernst effect in disordered superconducting films]{Nernst effect as a probe of superconducting fluctuations in disordered thin films}

\author{A Pourret\footnote{Present address:
James Franck Institute, University of Chicago, 5640 South Ellis
Avenue, Chicago, IL 60637, USA.}, P Spathis\footnote{Present
address: INAC, Grenoble, France.}, H Aubin and K Behnia}

\address{ Laboratoire Photons et Mati\`{e}re (CNRS), ESPCI, 10
rue Vauquelin, 75231 Paris, France} \ead{herve.aubin@espci.fr}

\begin{abstract}
In amorphous superconducting thin films of $Nb_{0.15}Si_{0.85}$ and
$InO_x$, a finite Nernst coefficient can be detected in a wide range
of temperature and magnetic field. Due to the negligible
contribution of normal quasi-particles, superconducting fluctuations
easily dominate the Nernst response in the entire range of study. In
the vicinity of the critical temperature and in the zero-field
limit, the magnitude of the signal is in quantitative agreement with
what is theoretically expected for the Gaussian fluctuations of the
superconducting order parameter. Even at higher temperatures and
finite magnetic field,  the Nernst coefficient is set by the size of
superconducting fluctuations. The Nernst coefficient emerges as a
direct probe of the ghost critical field, the normal-state mirror of
the upper critical field. Moreover, upon leaving the normal state
with fluctuating Cooper pairs, we show that the temperature
evolution of the Nernst coefficient is different whether the system
enters a vortex solid, a vortex liquid or a phase-fluctuating
superconducting regime.
\end{abstract}

\submitto{\NJP}
\maketitle
\section{Introduction}

Those past years have witnessed the emergence of the Nernst effect
as an important probe of superconducting fluctuations, following the
observation of an anomalous Nernst signal in the high temperature
phase of underdoped cuprates~\cite{Xu2000}. Because of the small
superfluid stiffness expected in underdoped
cuprates\cite{Emery1995}, and because vortices are a well-known
source of a Nernst response\cite{Vidal1973}, these authors proposed
the vortex-like excitations of a phase-disordered superconductor as
a natural source of this Nernst signal\cite{Wang2006}.

This discovery motivated numerous experimental and theoretical works
on the Nernst effect. On one hand, numerous studies on correlated
metals of various families resolved an unexpectedly large Nernst
coefficient (for a review see \cite{Behnia2008}). In some cases,
this was in total absence of superconductivity. The most
illuminating example was bismuth, the semi-metallic element in which
Nernst and Ettingshausen discovered in 1886 the effect which bears
their name\cite{Ettingshausen1886}. The Nernst coefficient in
bismuth\cite{Behnia2007} is three orders of magnitude larger than
what is typically seen in any type II superconductor. In fact, the
large magnitude of the Nernst coefficient in bismuth is in agreement
with the implications of the semiclassic transport
theory\cite{Oganesyan2004, Behnia2008,Varlamov2008} and therefore, a
large Nernst signal does not necessarily imply superconducting
fluctuations [either of phase or amplitude of the order parameter].

On the other hand, this led to the first theoretical study of the
Nernst response of fluctuating Cooper pairs\cite{Ussishkin2002}.
These fluctuations are usually described in the Gaussian
approximation within the Ginzburg-Landau framework\cite{LarkinBook}
and are known to give rise to the phenomena of
paraconductivity\cite{Glover1967}, i.e. an excess of conductance due
to short lived Cooper pairs in the normal state, and to the
so-called fluctuations diamagnetism\cite{Gollub1973}. Theoretical
calculations by Ussishkin, Sondhi and Huse (USH)\cite{Ussishkin2002}
have shown that Cooper pair fluctuations should also produce a
sizable Nernst signal, despite the absence of well defined
vortex-like excitations.

This prediction was put to test through measurements of the Nernst
effect in amorphous thin films of $low-T_c$ superconductors. The
normal state of these systems is a simple dirty metal with a totally
negligible Nernst response. These last
studies\cite{Pourret2006,Pourret2007,Spathis2008} demonstrated that
the Nernst signal of amorphous superconducting films is exclusively
generated by superconducting fluctuations, thus, providing a
remarkable testboard for theories. In quantitative agreement with
USH theory close to $T_c$, these measurements established that
conventional Gaussian fluctuations does indeed generate a Nernst
signal.

Following this observation, we now need to learn how to distinguish
other regimes of superconducting fluctuations from those simple
Cooper pair fluctuations, in particular, regimes with only thermal
or quantum fluctuations of the phase of the Superconducting Order
Parameter (SOP) as expected in the underdoped cuprates, or in the
vicinity of quantum superconductor-insulator transitions.
Furthermore, in presence of an applied magnetic field, we want to
learn how to distinguish the regime of Cooper pair fluctuations from
the vortex fluid with long-lived vortices that exist in any type-II
superconductor. Thus, one major ambition in the field is to identify
the characteristic signatures of those different regimes of
fluctuations in the Nernst data.

In this manuscript, we review our observation of the Nernst signal
by Cooper pair fluctuations and our identification of the Ghost
Critical Field (GCF) in the amorphous superconducting films
$Nb_xSi_{1-x}$\cite{Pourret2006,Pourret2007} and
$InO_x$\cite{Spathis2008}. Then we describe the evolution of the
Nernst signal within their superconducting phase diagram, from the
regime of Cooper pair fluctuations to the vortex solid, across the
vortex liquid. In finite magnetic field, a large increase in the
Nernst signal is  observed in the crossover from the regime of
Copper pair fluctuations to the vortex liquid phase, i.e. one
non-superconducting dissipative state. In the zero magnetic field
limit, where a true second order transition takes place between the
regime of Cooper pair fluctuations and the dissipationless vortex
solid, the Nernst coefficient diverges at the approach of the
superconducting transition, i.e. following the diverging correlation
length, and becomes zero in the vortex solid region. No abrupt
increase of the Nernst signal due to vortices is observed as the
temperature range for the existence of the vortex liquid shrinks to
zero in the zero magnetic field limit.

The organization of this paper is as follows. Section 2 describes
different regimes of superconducting fluctuations, whose existence
has been speculated in amorphous thin films or cuprates. Section 3
reviews samples characteristics and experimental setup. Section 4
describes the Nernst signal generated by the vortex flow; Section 5,
the Nernst signal generated by Cooper pair fluctuations. Section 6
describes the evolution of the Nernst coefficient across the
transition from the regime of Cooper pair fluctuations, i.e. normal
state, to the vortex solid. Finally, we discuss the effect of
thermal and quantum fluctuations of SOP on the Nernst response of
the vortex fluid.

\section{Regimes of superconducting fluctuations}

According to BCS theory, cooling a superconductor below its
superconducting transition temperature leads simultaneously to both
the Cooper pairs formation and their Bose condensation into a
macroscopically coherent quantum state. However, several subjects of
contemporary studies in superconductivity ask us to consider the
possibility that Cooper pairs may exist without macroscopic phase
coherence, mostly as a consequence of thermal or quantum
fluctuations of the
SOP\cite{Blatter1994,Fisher1991,Kosterlitz1973,Emery1995}. The
magnitude of these fluctuations and their predominance in the phase
diagram depends on materials parameters such as the amount of random
impurities, i.e. quenched disorder, dimensionality or correlation
length value\cite{Blatter1994}.

\begin{figure}
\begin{center}
{\includegraphics[width=15cm,keepaspectratio]{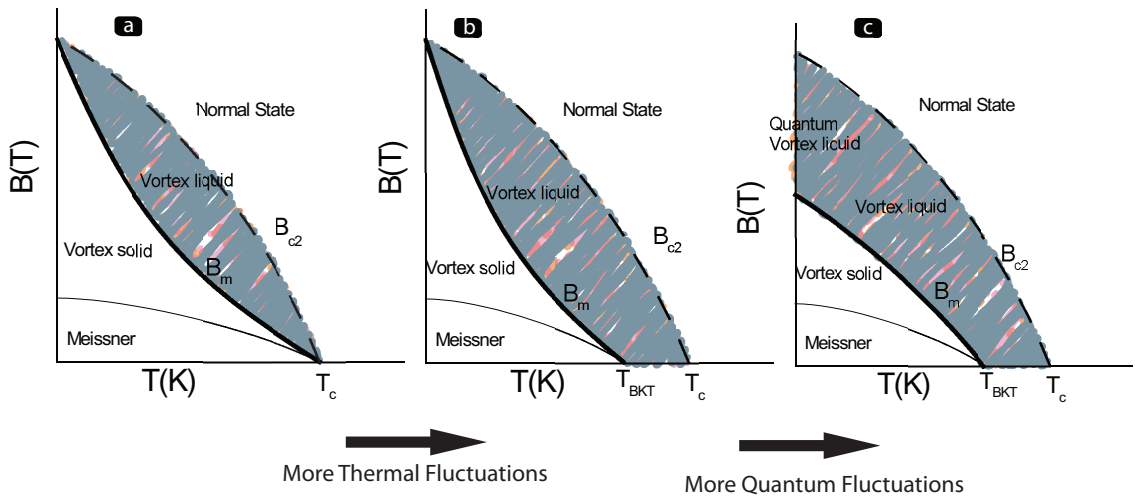}}
\caption{\label{fig:fig1} Evolution of the phase diagram of a
type-II superconductor as the effects of thermal fluctuations
increase-- panel a) to panel b)-- and the effects of quantum
fluctuations increase-- panel b) to panel c)--. A second order phase
transition, i.e. with diverging correlation length, separates the
vortex glass from the vortex liquid phase at $B_m$ (thick line).
Only a crossover is expected between the vortex liquid and the
normal state, at $B_{c2}$ (dashed line).}
\end{center}
\end{figure}

One such electronic phase is well known, found in many conventional
and non-conventional superconductors, the vortex-liquid phase. This
vortex fluid results from the melting of the vortex-solid above some
magnetic field scale $B_m$\cite{Fisher1991,Blatter1994}, as a
consequence of thermal fluctuations of the phase of SOP. This vortex
fluid is separated from the normal state only by a crossover at the
upper critical field $B_{c2}$, as shown on the phase diagram, panel
a) of figure~\ref{fig:fig1}.

In $high-T_c$ superconductors, a combination of high temperature,
small correlation length, large magnetic penetration depth and
quasi-two-dimensionality, conspire to increase the effects of
thermal fluctuations and $B_m$ can be significantly smaller than the
upper critical field $B_{c2}$.

In contrast, in bulk $low-T_c$ superconductor, $B_m$ almost
coincides with $B_{c2}$. However, as the vortex lattice is unstable
against the introduction of quenched disorder\cite{Larkin1979}, i.e.
random pinning sites, the superconducting phase diagram of amorphous
thin films usually displays a large vortex liquid region.

As the effects of thermal fluctuations are enhanced, either by
increasing disorder, reducing dimensionality, or reducing superfluid
density, a phase-disordered vortex liquid state may survive in the
limit of zero magnetic field \cite{Fisher1991,Blatter1994}, giving
rise to a phase diagram as shown panel b) of figure~\ref{fig:fig1}.
In this diagram, in the zero magnetic field limit, a second
temperature scale emerges for the establishment of
superconductivity, where macroscopic coherence sets in.

One similar situation has been intensively studied theoretically in
two dimensions by Berezinsky, Kosterlitz and Thouless
(BKT)\cite{Kosterlitz1973,Ambegaokar1980}. They found that, in two
dimensions and zero magnetic field, there exists a temperature scale
$T_{BKT}$ that correspond to a transition between two distinct
regimes of superconducting fluctuations where only the phase degree
of freedom are altered by the transition. The low temperature state
($T<T_{BKT}$) is quasi-ordered with algebraically decaying
correlation functions. The high temperature state ($T>T_{BKT}$) is
phase-disordered due to thermally generated vortex-antivortex pairs
that dissociate and populate the ground state. This leads to a
phase-incoherent superconducting state with exponentially decaying
correlation functions. Strict experimental realizations of this
model for charged superfluid is still lacking; however, some
variations of it are being considered to apply in some part of the
phase diagram of the cuprates and in the vicinity of the quantum
superconductor-insulator transition observed in amorphous and
granular superconducting thin films\cite{Goldman1998}.

In cuprates, the observation of a pseudo-gap above $T_c$, in the
underdoped region of their phase diagram, was interpreted as a
possible signature of two temperature scales for superconductivity.
The higher temperature scale, where the pseudo-gap forms in the
electronic spectrum, may correspond to Cooper-pairs formation, and
the second, lower temperature scale, akin to $T_{BKT}$, would
correspond to the transition toward the phase-coherent
superconducting state\cite{Lee2006}. This regime of phase-only
fluctuations is fundamentally different from the order parameter
fluctuations as described in the context of Ginzburg-Landau
theory\cite{LarkinBook}. In this last theory only one single
temperature scale, $T_c$, or magnetic field scale, $B_c$,
corresponding to the Cooper pair formation, is required to describe
the fluctuations. Remarkably, within the Ginzburg-Landau framework,
there is no upper temperature limit for the existence of these
fluctuations; they are expected to survive far above $T_c$ in the
normal state. In contrast, the regime of phase-only fluctuations
implies two distincts temperature or magnetic field scales: one
higher temperature scale for Cooper pair formation and one lower
temperature scale for the establishment of phase coherence. Between
these two temperatures, there exists a fluctuation regime
characterized by long-lived, phase-incoherent, Cooper pairs and
freely moving vortex-antivortex pairs. In the context of cuprates
physics, Emery and Kivelson\cite{Emery1995} extended the concept of
phase-coherence temperature introduced by BKT. They suggested that,
for any superconductor in any dimension, vortex-antivortex pairs
should appear spontaneously when the thermal energy $k_BT$ is larger
than the energy cost for their formation; this energy cost results
from the kinetic energy associated with superfluid flow around the
vortices. This defines a temperature scale for phase coherence,
$T_{COH}$, above which spontaneous nucleation of vortices is
possible. In conventional superconductors, this coherence
temperature largely exceeds $T_{BCS}$, the Cooper pair forming
temperature, and superconducting fluctuations exist only as
fluctuations of both the amplitude and phase of the SOP. In
contrast, for low density superfluid, as the underdoped cuprates,
$T_{COH}<T_{BCS}$. This implies that the temperature for the
superconducting transition is controlled by the superfluid density.
In the context of cuprates physics, this provides an explanation of
the Uemura plot\cite{Uemura1989} where $T_c$ is found to scale with
the magnetic penetration depth, which is inversely proportional to
superfluid density.

Finally, in addition to quenched disorder and thermal fluctuations,
quantum fluctuations of the SOP provides another origin for the
\emph{quantum} melting of the vortex solid. This leads to a phase
diagram as shown panel c) of figure~\ref{fig:fig1}, where a quantum
liquid of vortices is expected in the zero-temperature limit,
separated from the superconducting state by a second order
transition whose critical behavior is controlled by quantum
fluctuations~\cite{Sondhi1997}. Fine-tuning of the transition can be
achieved either by applying a perpendicular magnetic
field~\cite{Hebard1990,Paalanen1992,Yazdani1995,Ephron1996,Markovic1998a,Markovic1998b,Gantmakher2000,
Bielejec2002,Sambandamurthy2004,Aubin2006} or by varying the sheet
resistance $R_{\square}$ of the films -- using film
thickness\cite{Jaeger1989,Markovic1999,Kikuchi2008} or electrostatic
field\cite{Parendo2005}.

The systems discussed in this manuscript are amorphous
superconducting films for which distinct regimes of superconducting
fluctuations are possible. Well above the mean field superconducting
transition $T_c$, we expect the conventional Cooper pair
fluctuations; below $T_c$, different regimes may exists according to
the amount of thermal or quantum phase fluctuations. One quantum
origin is possible as quantum superconductor-insulator transitions
have been observed in both
systems\cite{Sambandamurthy2004,Aubin2006}.

\section{The compounds studied and the experimental technique used}

In this paper we present the evolution of the Nernst signal across
the phase diagram of two different disordered superconductors,
$Nb_{0.15}Si_{0.85}$ and $InO_x$.

The two amorphous thin films of Nb$_{0.15}$Si$_{0.85}$ used for this
study were prepared by L. Dumoulin's group. The samples are
deposited by co-evaporation of Nb and Si in an ultra-high vacuum
chamber, as described elsewhere\cite{Dumoulin1993,marnieros2000}. On
the other hand, the 300 \AA{}-thick amorphous InO$_x$ film was
prepared by Z. Ovadyahu's group. The sample is deposited on a glass
substrate by $e$-gun evaporation of In$_2$O$_3$ in oxygen
atmosphere~\cite{Ovadyahu1993}. The as-prepared film has an
insulating-like behavior down to the lowest measured temperature of
60~mK. After thermal annealing at $50^\circ$C under vacuum as
described elsewhere~\cite{Ovadyahu1986}, the room temperature sheet
resistance decreases by about 30 \% and a superconducting state
appears. During all measurements, the film has been kept below
liquid nitrogen temperature to avoid aging effects.

 Several characteristics of
$InO_x$ indicate that effects of thermal or quantum fluctuations are
stronger in this system than in $Nb_{0.15}Si_{0.85}$. While
$Nb_{0.15}Si_{0.85}$ has a high carrier density $n=8.10^{22}
cm^{-3}$, comparable to any ordinary metal, the carrier density of
$InO_x$ is 80 times smaller, $n=10^{21} cm^{-3}$, comparable to
values found for the underdoped cuprates. According to an argument
put forward by Emery and Kivelson\cite{Emery1995}, this low carrier
density increases the probability for the spontaneous nucleation of
vortices and so the amount of phase fluctuations. A second
difference between both systems is the larger sheet resistance of
$InO_x$, $R_\square \approx 4000\Omega$, which implies enhanced
quantum fluctuations with respect to $Nb_{0.15}Si_{0.85}$,
$R_\square \approx 350\Omega$. Finally, one last striking difference
between both system is the observation of a large negative
magnetoresistance in $InO_x$. This phenomena has been interpreted as
a possible indication of the pair-breaking effect of magnetic field
on localized Cooper
pairs\cite{Gantmakher2000,Gantmakher2001,Steiner2005}

The Nernst effect is the transverse thermoelectric response
$N=E_y/\nabla_x T$ of a sample submitted to a thermal gradient and a
magnetic field applied perpendicular to sample plane. One usually
defines the Nernst coefficient $\nu=N/B$, and within linear response
theory, one also defines the Peltier conductivity tensor:

\begin{equation}
\label{tensor}
    \left(
    \begin{array}{c}
    \mathbf{j}_e\\
    \mathbf{j}_{th}
    \end{array}
    \right)
    =
    \left(\begin{array}{cc}
    \hat{\sigma}&\hat{\alpha}\\
    \hat{\tilde{\alpha}}&\hat{\kappa}
    \end{array}\right)
    \left(\begin{array}{c}
    \mathbf{E} \\
    \nabla \mathbf{T}
    \end{array}\right)
\end{equation}

From the condition, $\mathbf{j}_e=0$, one gets:

\begin{equation}
N=\frac{\sigma_{xx}\alpha_{xy}-\sigma_{xy}\alpha_{xx}}{\sigma_{xx}^2+\sigma_{xy}^2}
\end{equation}

For all samples discussed, the Hall angle is small, and so is
$\sigma_{xy}$. This leads to a simple relationship between the
Nernst coefficient $\nu$ and the Peltier coefficient $\alpha_{xy}$.
\begin{equation}
\nu\approx\frac{\alpha_{xy}}{B\sigma_{xx}}
\end{equation}

In our experimental setup, the Nernst signal is measured using a one
heater - two $RuO_2$ thermometers setup. It allows measurements of
diagonal and off-diagonal thermoelectric and electric transport
coefficients with the same contacts. At low temperature, $T<4.2K$, a
DC voltage of $1 nV$ can be resolved and typical relative resolution
of $10^{-3}$ on the magnitude of temperature gradient can be
achieved.

In superconductors, the two most important contributions expected
are, below $T_c$, the vortex contribution, $N^S$, and above $T_c$,
the normal electrons contribution, $N^n$. The measured Nernst signal
is the sum of both contributions.
\begin{equation}
N=N^S+N^n
\end{equation}

In the amorphous superconductors studied here, the Nernst signal due
to normal quasiparticles is particularly low as this contribution
scales with electron mobility\cite{Behnia2008}. This characteristic
of amorphous superconductors is of the utmost importance as it
allows to identify unambiguously the Nernst signal measured deeply
into the normal state with the contribution of superconducting
fluctuations.

Part of the Nernst data presented here have been discussed
previously, where we have shown that, in $Nb_{0.15}Si_{0.85}$,
Cooper pair fluctuations could generate a Nernst signal up to very
high temperature ($30 \times T_c$) and high magnetic field ($4
\times B_{c2}$) in the normal state\cite{Pourret2006,Pourret2007}.
In this regime, we found that the magnitude of the Nernst
coefficient is set by the size of superconducting fluctuations and
led to emergence of a field scale above $T_c$, the Ghost Critical
Field (GCF), whose value is set by the correlation
length\cite{Pourret2007}. Tracking the temperature dependence of the
GCF in $Nb_{0.15}Si_{0.85}$ and $InO_x$ demonstrates that both
systems have characteristically distinct behaviors across the
transition. In $Nb_{0.15}Si_{0.85}$, a true superconducting
transition is observed, while $InO_x$ is characterized by a large
region of superconducting fluctuations that prevent the
establishment of the superconducting order\cite{Spathis2008}.

\section{Long-lived vortices and Nernst effect}

Previous works on conventional
superconductors\cite{Huebener1969,Vidal1973} and
cuprates\cite{Ri1994,Wang2006} have shown that a large Nernst signal
is generated by vortices as they are displaced by an applied heat
current. This can be described phenomenologically by considering the
forces exerted on the vortices. There is the force exerted by the
thermal gradient, $\mathbf{f}=S_\phi(-\nabla T)$ where $S_\phi$ is
the entropy transported per vortex. Moving vortices with speed
$\mathbf{v}$ are also subject to the frictional force
$\mathbf{f}_f=\eta\mathbf{v}$, where the damping viscosity $\eta$
may be inferred from the the flux-flow resistivity
$\rho=B\phi_0/\eta$ where $\phi_0=h/2e$ is the superconducting flux
quantum. In the steady state, when the frictional force balances the
thermal force, the Nernst signal is given by :

\begin{equation}
N=\frac{Bs_\phi}{\eta}=\frac{\rho s_\phi}{\phi_0}
\end{equation}

\begin{figure}
\begin{center}
{\includegraphics[scale=1]{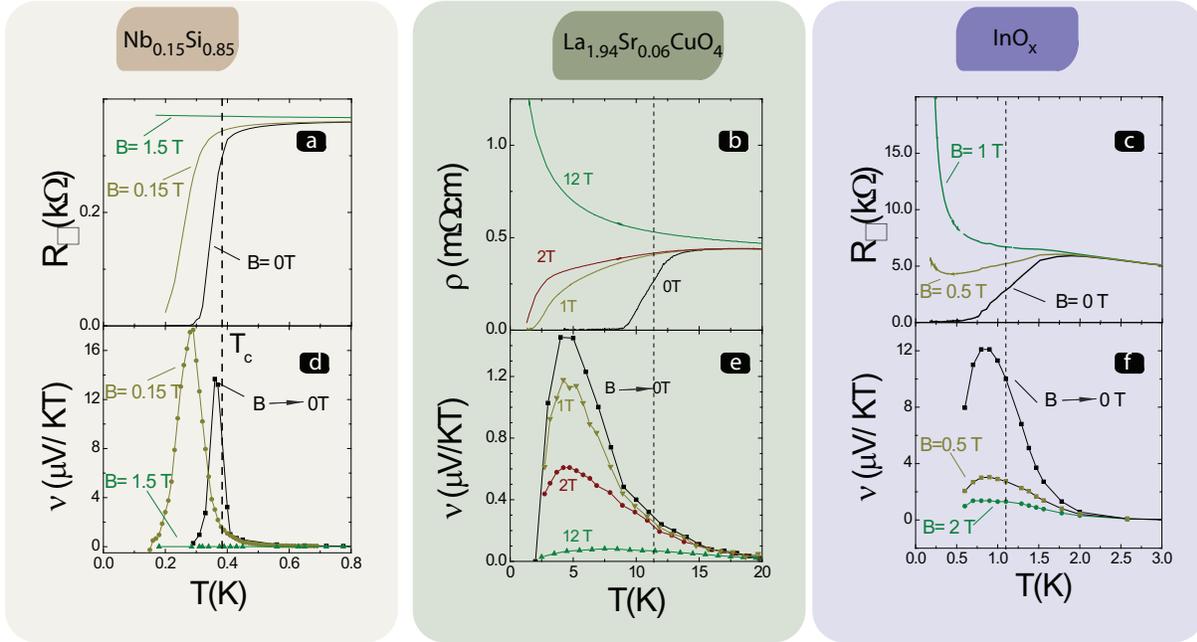}}
\caption{\label{fig:fig2} Sheet resistance, panels a) and c),
resistivity, panel b) and Nernst data shown in panels d), e) and f)
as a function of temperature for $Nb_{0.15}Si_{0.85}$,
$La_{1.94}Sr_{0.06}CuO_4$ and $InO_x$.}
\end{center}
\end{figure}

Figure~\ref{fig:fig2} shows the temperature dependence of
resistivity and Nernst coefficient data across the superconducting
transition of one $35 nm$ thick film of $Nb_{0.15}Si_{0.85}$, one
$30 nm$ thick film of $InO_x$ and the undedoped cuprate
$La_{1.94}Sr_{0.06}CuO_4$, taken from references
\cite{Pourret2006,Pourret2007}, \cite{Spathis2008} and
\cite{Capan2003} respectively. For $Nb_{0.15}Si_{0.85}$, we observe
a sharp increase of the Nernst coefficient at the superconducting
transition. As we will see later, in the zero magnetic field limit,
this large enhancement of the Nernst coefficient reflects the
diverging correlation length at the approach of the superconducting
transition. While the Nernst signal due to superconducting
fluctuations appears simply as the high temperature tail of the
large vortex-induced Nernst signal observed below $T_c$. We will
show that a comparison of the magnetic field dependence of the
Nernst signal, figure~\ref{fig:fig3}, measured above and below
$T_c$, allows to establish a fundamental distinction between the
data measured above and below $T_c$. At finite magnetic field, as
the only genuine superconducting phase is the dissipation-less
vortex solid, the large enhancement of the Nernst coefficient
actually reflects a crossover between two regimes of fluctuations,
the regime of Cooper pair fluctuations and the vortex fluid with
frozen amplitude fluctuations of the order parameter.

For $InO_x$ and $La_{1.94}Sr_{0.06}CuO_4$ , figure~\ref{fig:fig2}
shows that the Nernst coefficient changes continuously across the
transition and does not increase abruptly at the transition. For
$InO_x$, this reflects the absence of a true phase transition, with
diverging correlation length, and so the absence of long range
superconducting order in this system.

\section{Cooper pair fluctuations and Ghost critical field}

\begin{figure}
\begin{center}
{\includegraphics[width=10cm,keepaspectratio]{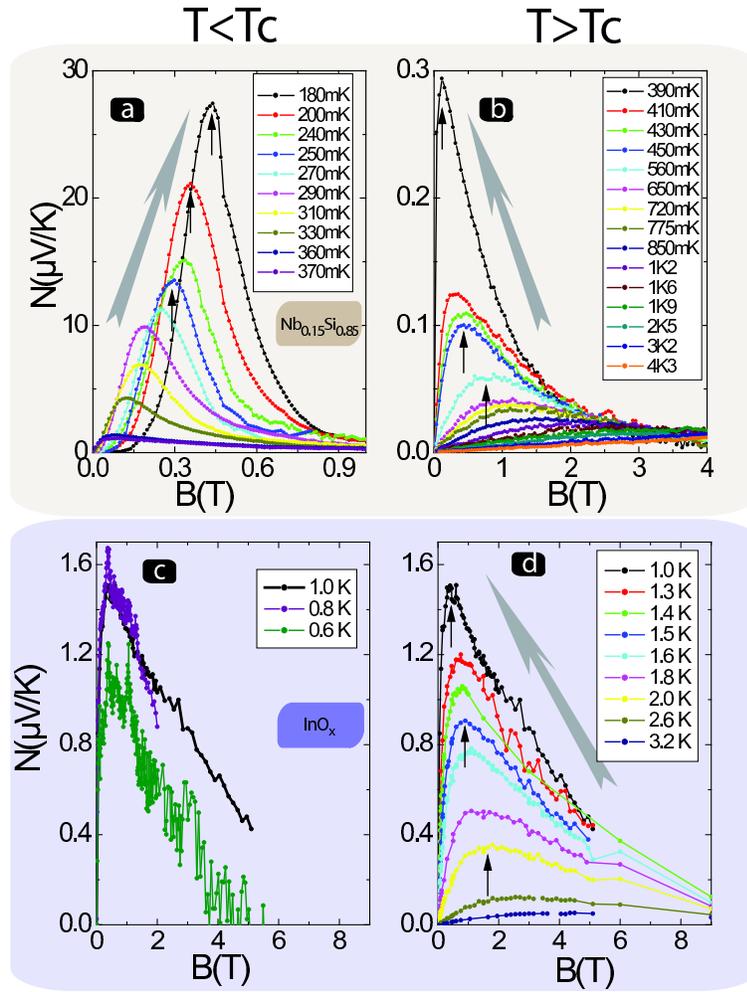}}
\caption{\label{fig:fig3} Nernst signal measured below and above
$T_c$: for $Nb_{0.15}Si_{0.85}$, panel a) and panel b), respectively
and for $InO_x$, panel c) and panel d), respectively. The maxima
occurring at $B^*$ are indicated by arrows. Below $T_c$, $B^*$
increases toward low temperature, like $B_{c2}$ and $B_m$. Above
$T_c$, the temperature dependence of $B^*$ is reverted, it increases
with increasing temperature as expected for the GCF.}
\end{center}
\end{figure}

Figure~\ref{fig:fig3} shows the magnetic field dependence of the
Nernst signal for $Nb_{0.15}Si_{0.85}$ and $InO_x$. In the normal
state, for both systems, the Nernst data follow a characteristic
tilted tent profile with a maximum at the field scale $B^*$ whose
magnitude is observed to increase with temperature.

\begin{figure}
\begin{center}
{\includegraphics[width=15cm,keepaspectratio]{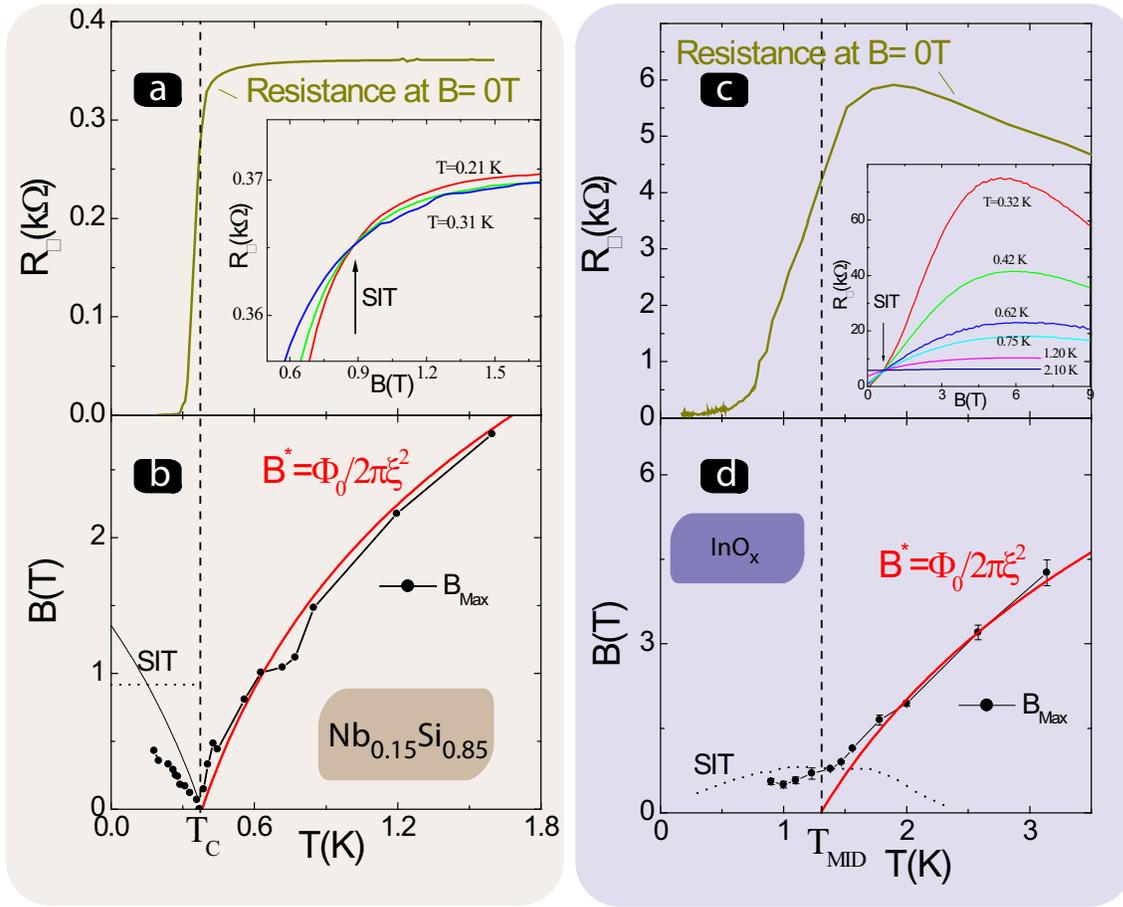}}
\caption{\label{fig:fig4} Top panels: Resistance curves of
$Nb_{0.15}Si_{0.85}$, panel a), and $InO_x$, panel c). Bottom
panels: Phase diagram displaying the field scale $B^*$ as function
of temperature. For $Nb_{0.15}Si_{0.85}$, panel b), this field scale
goes to zero at $T_c$. Below $T_c$, this field scale reflects the
field position where the vortex-induced Nernst signal reaches a
maximum. Above $T_c$, this field scale reflects the GCF. For
$InO_x$, panel d), only the GCF is clearly identified in the data.
It keeps decreasing as the temperature is swept across the
superconducting transition. In contrast to $Nb_{0.15}Si_{0.85}$,
there is no distinct signature of the large Nernst signal due to
vortex flow. For both samples is also shown the critical field for
the SIT as extracted from crossing point of the resistance curves
plotted as function of magnetic field, insets of top panels.}
\end{center}
\end{figure}

Below $T_c$, for $Nb_{0.15}Si_{0.85}$, the vortex-induced Nernst
signal increases steeply with magnetic field, when the vortices
become mobile following the melting of the vortex solid state. It
reaches a maximum and decreases at larger magnetic fields when the
excess entropy of the vortex core is reduced. In contrast to the
high temperature regime, the position of the maximum $B^*$ shifts
toward higher magnetic fields upon decreasing the temperature. This
is not surprising, since in the superconducting state, all field
scales associated with superconductivity, as $B_{c2}$ and $B_m$, are
expected to increase with decreasing temperature. Plotting the
position of $B^*$, above and below $T_c$, on the phase diagram
figure~\ref{fig:fig4} shows that $B^*$ goes to zero just at $T_c$.
This observation is the most definitive signature that the
superconducting fluctuations at the origin of the Nernst signal
observed above $T_c$ are of a fundamentally distinct nature than
below $T_c$. Below $T_c$, the Nernst signal is generated by the
long-lived vortices of the vortex fluid, above $T_c$, the Nernst
signal is generated by Cooper pair fluctuations.

These fluctuations correspond to spatial and temporal fluctuations
of the SOP $\Psi(x,t)$ and are described by the Ginzburg-Landau
theory\cite{LarkinBook}. The typical size of these superconducting
fluctuations is set by the correlation length $\xi_d$. It
characterizes the length scale on which the correlation function
$<\psi(x_0)\psi(x_0-x)>$ decreases to zero. Upon cooling, this
correlation length increases and diverges at the approach of the
superconducting transition as $\xi_d=\xi_0 \epsilon^{-1/2}$ where
$\epsilon=ln(T/T_c)$ is the reduced temperature. At the microscopic
level, these fluctuations correspond to short-lived Cooper pairs
whose life-time is controlled by their decay into free electrons :

\begin{equation}
\tau=\frac{\pi\hbar}{8k_BT_c}\epsilon^{-1}
\end{equation}

These Cooper pairs fluctuations give rise to the phenomena of
paraconductivity\cite{Glover1967} and fluctuation
diamagnetism\cite{Gollub1973}. As normal quasiparticles contribute
significantly to conductivity and magnetic susceptibility, the
sensitivity of these probes to superconducting fluctuations is
limited to a narrow region close to the superconducting
transition\cite{Skocpol1975}. In contrast, in these amorphous films,
as the elastic mean free path is only a few Angstrom long, the
contribution of free electron to the Nernst signal is particularly
weak, orders of magnitude lower than the measured Nernst signal due
to superconducting fluctuations. This explains that the Nernst
signal generated by short-lived Cooper pairs can be detected up to
very high temperatures ($30\times T_c$) and high magnetic field
($4\times B_{c2}$), deep into the normal
state\cite{Pourret2006,Pourret2007}. Furthermore, because of this
weak contribution of normal quasiparticles excitations, a direct and
unambiguous comparison of Nernst data with superconducting
fluctuations theories is possible.

\begin{figure}
\begin{center}
{\includegraphics[width=15cm,keepaspectratio]{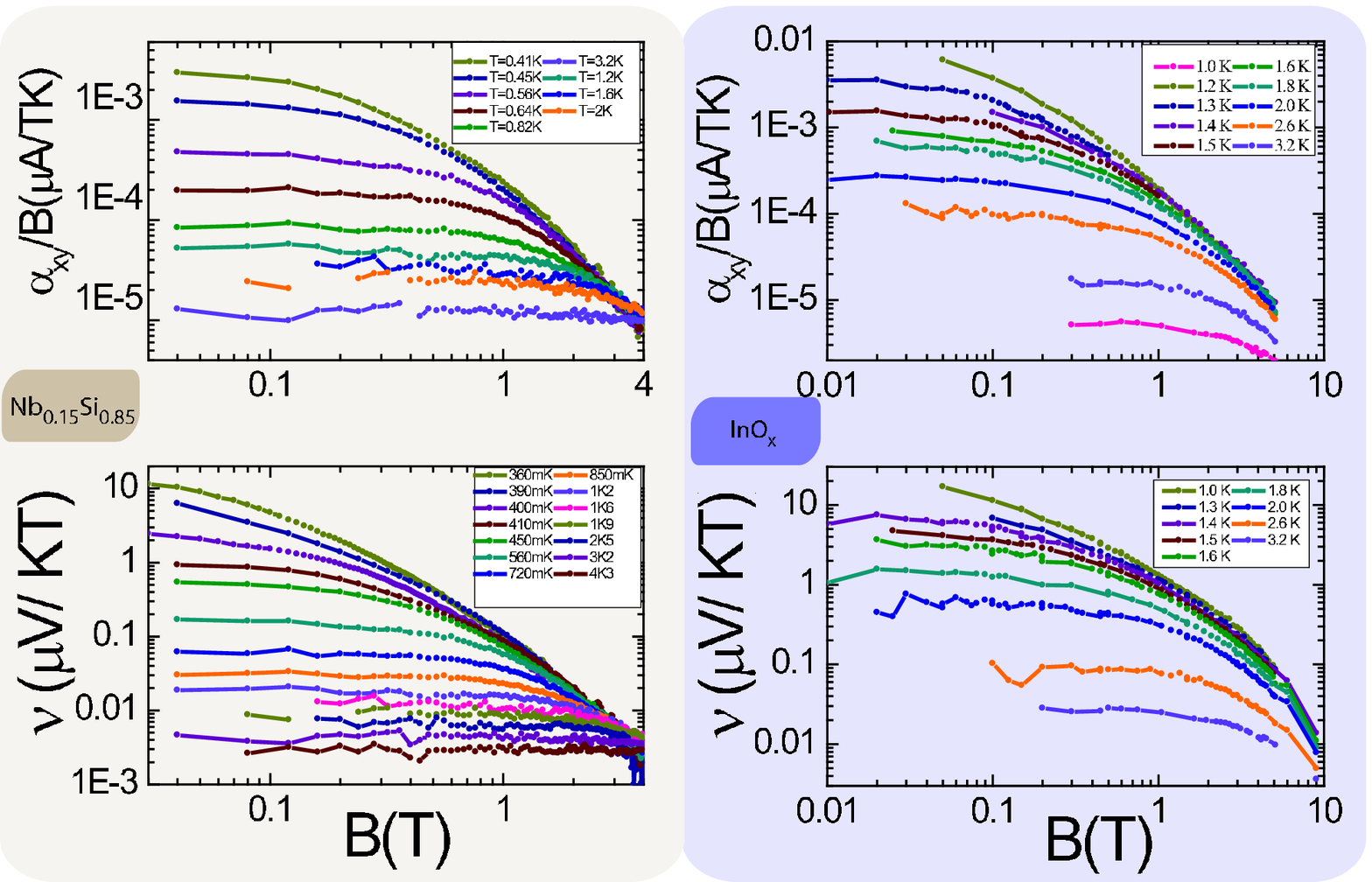}}
\caption{\label{fig:fig5} Nernst coefficient $\nu$ (bottom panels)
and Peltier coefficient  $\frac{\alpha^{SC}_{xy}}{B}$ (top panels)
for $Nb_{0.15}Si_{0.85}$ (left) and $InO_x$ (right). The similarity
between the plots show that the evolution of the Peltier coefficient
is controlled by the variations of the Nernst coefficient. For both
systems, we find that at low field $B<B^*$ those coefficients are
independent of magnetic field, they are set only by the temperature
dependent correlation length. In the opposite limit, $B>B^*$, those
coefficients are independent of temperature, they are determined by
the magnetic length.}
\end{center}
\end{figure}

Treating the fluctuations of the SOP in the Gaussian approximation,
USH obtained a simple analytical formula, valid close to $T_c$ and
in the zero-magnetic field limit, relating the off-diagonal Peltier
coefficient $\alpha_{xy}$ to fundamental constants and the
correlation length\cite{Ussishkin2002}.

\begin{equation}\label{eq:eq1}
   \frac{\alpha^{SC}_{xy}}{B}= \frac{1}{6\pi}\frac{k_{B} e^2}{\hbar^2}\xi^{2}
\end{equation}

where $\frac{\alpha_{xy}}{B}$ is simply related to the Nernst
coefficient and the measured conductivity through the formula
$\frac{\alpha_{xy}}{B}\approx\sigma_{xx}\nu$. Above $T_c$, as the
conductivity of samples change only weakly with temperature and
magnetic field, the evolution of the Peltier coefficient is mostly
controlled by the Nernst coefficient value, as shown
figure~\ref{fig:fig5} where $\nu$ and $\frac{\alpha_xy}{B}$ are
plotted side by side.

One remarkable characteristic of formula~\ref{eq:eq1} is that the
coefficient $\alpha^{SC}_{xy}/B$ is independent of magnetic field. A
plot of this coefficient obtained experimentally for
$Nb_{0.15}Si_{0.85}$ and $InO_x$, Figure~\ref{fig:fig5}, shows that
this is indeed the case at low magnetic field.

\begin{figure}
\begin{center}
{\includegraphics[width=15cm,keepaspectratio]{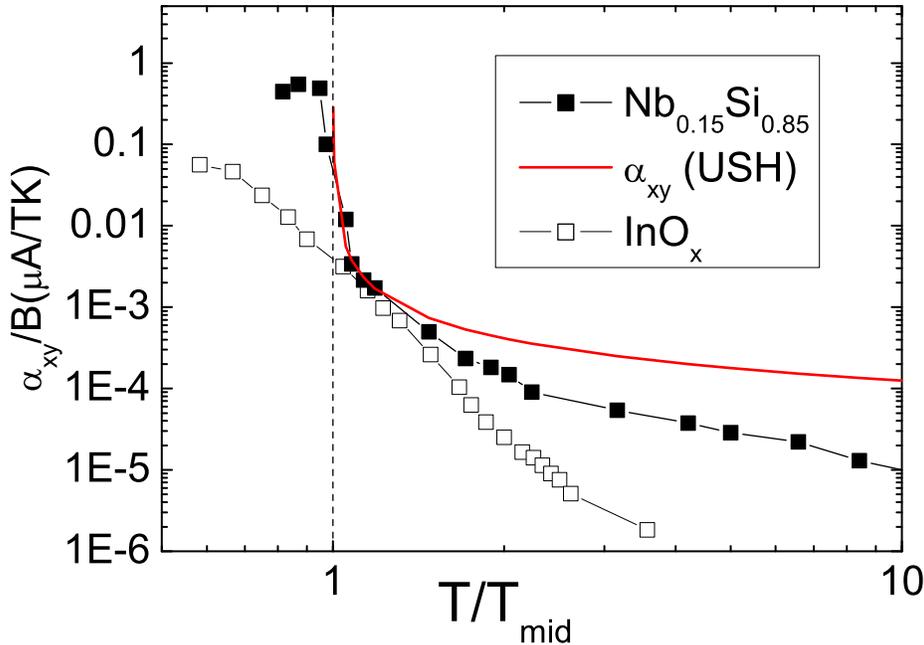}}
\caption{\label{fig:fig6} Peltier coefficient
$\frac{\alpha^{SC}_{xy}}{B}$ in the zero magnetic field limit
plotted as function of temperature for $Nb_{0.15}Si_{0.85}$ and
$InO_x$. The data of $Nb_{0.15}Si_{0.85}$ are compared with USH
theory.}
\end{center}
\end{figure}

From those plots, the value of $\frac{\alpha^{SC}_{xy}}{B}$ in the
zero magnetic field limit, $(B\rightarrow 0)$, is extracted and
compared to USH equation~\ref{eq:eq1}, as shown
figure~\ref{fig:fig6}.

For $Nb_{0.15}Si_{0.85}$, a quantitative agreement with a
theoretical prediction is found close to $T_c$. At high temperature,
the data deviates from the USH theoretical expression. Recent
theoretical works have extended the calculations of the Nernst
effect due to Gaussian fluctuations beyond the regime of validity of
USH theory, to higher temperature and magnetic
field\cite{Serbyn2008,Michaeli2008} and have been found to be in
quantitative agreement with those data as well.

Thus, these last experimental and theoretical works have established
that well defined vortex-like excitations are \emph{not} required
for superconducting fluctuations to generate a Nernst signal, and
that the magnitude of the Nernst coefficient in the regime of
Gaussian fluctuations is simply related to the correlation length.
Remarkably, these measurements also demonstrated that even at high
magnetic field and high temperature, the Nernst coefficient is
simply related to that single length scale, the size of
superconducting fluctuations~\cite{Pourret2007,Spathis2008}. In the
zero-field limit, this size is set by the correlation length
$\xi_d$. In the high field limit, the size of superconducting
fluctuations is set by the magnetic length
$\ell_B=(\hbar/2eB)^{1/2}$ when this length becomes shorter than the
correlation length at zero magnetic field.

The shrinking effect of the magnetic field on superconducting
fluctuations is well known from studies of fluctuations diamagnetism
in low temperature superconductors\cite{Gollub1973} and
cuprates\cite{Carballeira2000}. While in the low field limit, the
magnetic susceptibility should be independent of the magnetic field
-- i.e. in the Schmidt limit\cite{Schmid1969} --, the magnetic
susceptibility is experimentally observed to decrease with the
magnetic field, following the Prange's formula\cite{Prange1970};
which is an exact result within the Ginzburg-Landau formalism. At
high magnetic field, the superconducting fluctuations are described
as evanescent Cooper pairs arising from free electrons with
quantized cyclotron orbits\cite{Skocpol1975}.

As a consequence of this phenomena, at a given temperature $T>T_c$,
the size of superconducting fluctuations decreases from the value
$\xi_d(T)=\xi_0 \epsilon^{-1/2}$, at low magnetic field, to the
magnetic length value $\ell_B$, when the magnetic field exceeds
$B^*=\phi_0/2\pi{\xi_d}^2$. This field scale was identified the
first time by Kapitulnik \emph{et al.} in the magnetoresistance data
of mixture films of InGe\cite{Kapitulnik1985}. As it mirrors, above
$T_c$, the upper critical field below $T_c$, it has been dubbed the
"Ghost critical Field", by these last authors.

As shown in figure~\ref{fig:fig3}, above $T_c$, this crossover is
responsible for the observed maximum in the field dependence of the
Nernst signal. Upon increasing the magnetic field, the Nernst signal
increases linearly with field, reaches a maximum at $B^*$ and
decreases beyond that field scale. As extensively discussed in our
previous publications\cite{Pourret2007,Spathis2008}, we recall here
the arguments demonstrating that the Nernst coefficient is set by
the size of superconducting fluctuations and that $B^*$ is set by
the GCF.

\begin{figure}
\begin{center}
{\includegraphics[width=10cm,keepaspectratio]{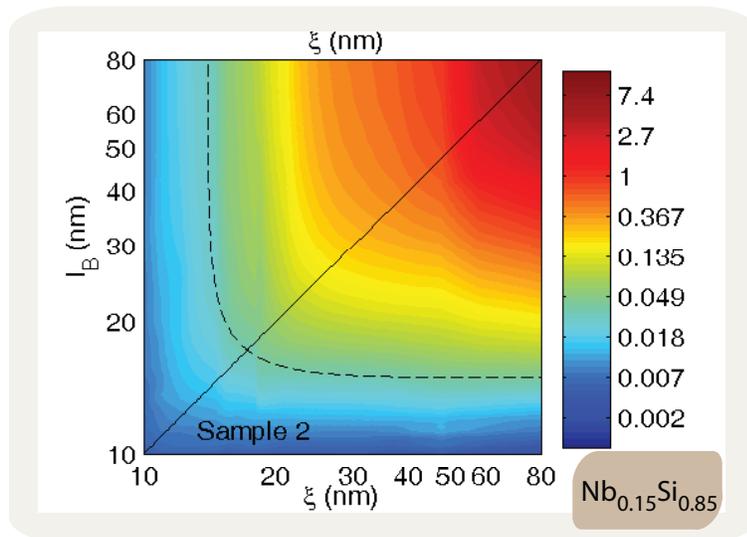}}
\caption{\label{fig:fig7} Logarithmic color map of the Nernst
coefficient as a function of the magnetic length $\ell_B$ and the
zero-field correlation length $\xi_d$ for $Nb_{0.15}Si_{0.85}$. Note
the symmetry with respect to the diagonal continuous line
($\ell_{B}=\xi_d$).}
\end{center}
\end{figure}

\begin{itemize}
    \item At low magnetic field, the Nernst coefficient depends only on
the temperature and is independent of the magnetic field. Indeed,
when $\ell_B>\xi_d$, the size of the superconducting fluctuations is
set by the temperature dependent correlation length $\xi_d$. See
figure~\ref{fig:fig5}.
    \item Above $T_c$, the magnitude and the temperature dependence of $B^*$ follows the
field scale set by the Ginzburg-Landau correlation length
$\xi_{d}=\frac{\xi_{0d}}{\sqrt{\epsilon}}$ through the relation
$B^{*}=\frac{\phi_{0}}{2\pi \xi_{d}^2}$ where $\phi_{0}$ is the flux
quantum and $\epsilon=\ln{\frac{T}{T_{c}}}$ the reduced temperature.
See \cite{Pourret2006} and \cite{Spathis2008} for the details
regarding the determination of the correlation length in
$Nb_xSi_{1-x}$ and $InO_x$ respectively. The position of the maximum
$B^*$ is the field scale where $\ell_B=\xi_d$. As shown in the panel
b of figure~\ref{fig:fig4} for $Nb_{0.15}Si_{0.85}$, it mirrors
above $T_c$, the upper critical field below $T_c$.

    \item At high magnetic field, $B>B^*(T)$, the data for Nernst coefficient converge toward a
     weakly temperature-dependent curve. Indeed, when $\ell_B<\xi_d$, the
     size of superconducting fluctuations is set by the magnetic
     length, which is obviously independent of temperature. See
figure~\ref{fig:fig5}.
    \item As shown figure~\ref{fig:fig7} for $Nb_{0.15}Si_{0.85}$, when one
    substitutes temperature and magnetic field by their associated
    length scales: the zero-field superconducting correlation length
    $\xi_d(T)$ and the magnetic length $\ell_B(B)$, we find that the Nernst coefficient is symmetric with
    respect to the diagonal $\xi_d(T)=\ell_B$. This shows that the
    Nernst coefficient depends only on the size of superconducting
    fluctuations, no matter what sets it, the magnetic length or the
    correlation length.
\end{itemize}

Finally, we noticed previously for $Nb_{0.15}Si_{0.85}$ that $B^{*}$
goes to zero at $T_c$. It appears now clearly that this is the
consequence of the divergence of the correlation length at the
transition, which drives the GCF to zero. This characteristic
temperature dependence of $B^{*}$ is a remarkable signature of the
superconducting transition and is expected in any conventional
superconductor with a phase diagram as depicted in the panel a of
figure~\ref{fig:fig2}.

A quite distinct phenomena is observed in $InO_x$. $B^{*}$ keeps
decreasing on the temperature range where the superconducting
transition is expected, according to resistivity measurements. This
indicates that the correlation length does not diverge in this
sample, implying the absence of a true superconducting transition.
Most likely, strong superconducting fluctuations prevent the
establishment of the superconducting order in this
sample\cite{Spathis2008}. These fluctuations could also be hold
responsible for the weak vortex-induced Nernst signal in this
system. Indeed, the nature of vortices existing in conventional
vortex fluids is quite distinct from the vortex-like excitations
expected in BKT-type fluctuating regime. While vortices are
long-lived in the vortex fluid, they have a short life-time in
presence of phase fluctuations of the SOP. Most likely, such a
reduction of the life-time of vortices should reduce the Nernst
signal.

This situation bears much similarity with the underdoped cuprates,
where the weak Nernst signal observed at high temperature has been
attributed to short-lived vortex excitations of a regime with
phase-only superconducting fluctuations. However, in contrast to our
$InO_x$ sample, where the superconducting order is never reached in
our measurements, a genuine superconducting transition, with
diverging correlation length, occurs in the cuprates. Consequently,
as for $Nb_{0.15}Si_{0.85}$, it is expected that the GCF should
decrease to zero at $T_c$. While this field scale has never been
discussed and identified in the magnetic field dependence of the
Nernst signal in cuprates, it appears clearly in the Nernst data
shown figures 11, 12, 15, 16 from\cite{Wang2006} for
$Bi_2Sr_{1.6}La_{0.4}CuO_6$, $Bi_2Sr_{1.8}La_{0.2}CuO_6$,
$La_{1.83}Sr_{0.17}CuO_4$ and $Bi_2Sr_{1.6}La{0.4}CuO_6$,
respectively.

Despite the distinct characteristics of the three family of
materials discussed, $Nb_{0.15}Si_{0.85}$, $InO_x$ and the cuprates,
we find that the GCF is a robust feature of the Nernst signal
generated by superconducting fluctuations, no matter the precise
nature of those fluctuations, i.e. Cooper pair fluctuations or
phase-only fluctuations of the SOP. As a measure of the temperature
dependence of the correlation length, the GCF provides a remarkable
tool for the characterization of superconducting fluctuations.

\section{From Cooper pair fluctuations to the vortex fluid  }

As discussed earlier, $B_m$, the melting field of the vortex solid
is believed to be the only second order transition within the
temperature-magnetic field phase diagram of disordered type-II
superconductors. On the other hand, the upper critical field line
$B_{c2}$ is believed to represent only a crossover between the
vortex fluid and the regime of Cooper pair fluctuations. As we
established that, in the zero magnetic field limit, the Nernst
coefficient diverges at the transition as the correlation length,
this led us to speculate that the evolution of the Nernst
coefficient across the superconducting transition should be markedly
different at finite magnetic field. Indeed, while in the zero field
limit, the transition occurs directly between the regime of Cooper
pairs fluctuations and the vortex solid; at finite magnetic field,
the vortex fluid emerges between those two phases and prevents the
divergence of the correlation length within the regime of Cooper
pair fluctuations.

\begin{figure}
\begin{center}
{\includegraphics[width=10cm,keepaspectratio]{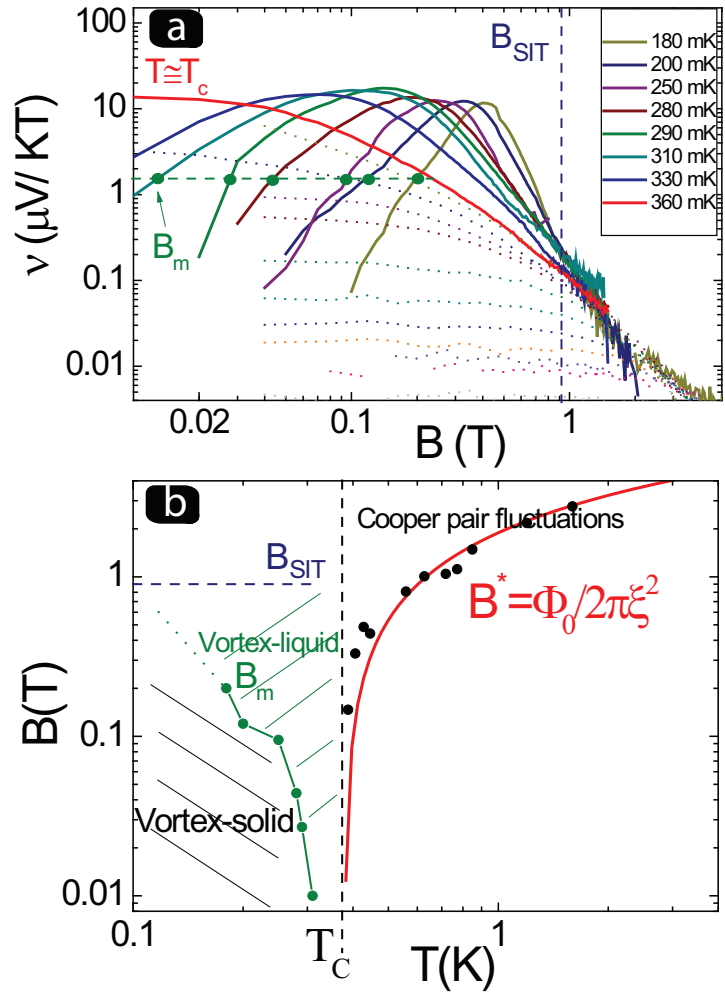}}
\caption{\label{fig:fig8} Panel a): Magnetic field dependence of
Nernst coefficient of $Nb_{0.15}Si_{0.85}$ for temperatures above
$T_c$ (dotted lines) and below $T_c$ (continuous lines). Panel b):
Phase diagram of $Nb_{0.15}Si_{0.85}$ on a log scale. See text for
the determination of three field scales: the GCF $B^*$, the SIT
critical field $B_{SIT}$ and the melting field $B_m$ of the vortex
solid.}
\end{center}
\end{figure}

To locate the vortex fluid within the phase diagram of
$Nb_{0.15}Si_{0.85}$, figure~\ref{fig:fig8}, panel a) shows the
Nernst coefficient as function of magnetic field measured at
temperatures above and below $T_c$.

The high field boundary of the vortex fluid phase is defined as the
field scale below which the Nernst signal exceeds values expected
for Cooper pair fluctuations. On this figure, we  see that the curve
at $T_c$ provides an upper envelop for the Nernst curves measured
above $T_c$ (the dotted lines) and a separatrix with the curves
measured below $T_c$. All these curves merge with the curve measured
at $T_c$ above a field scale about $0.9 T$. This field scale turns
out to  be close to the critical field $B_{SIT}$ for the
superconductor-insulator transition. This transition is identified
through the observation of a crossing point in the field dependence
of resistivity curves, as shown in insets of figure~\ref{fig:fig4},
and finite size scaling of the data\cite{Aubin2006}. Our
measurements show that the vortex-induced Nernst signal may be
damped by this transition. This is an unexpected observation as the
usual understanding of the superconductor-insulator transition
implies that the insulating phase should correspond to a quantum
fluid of vortices.

The low field boundary of the vortex fluid phase is obtained as the
field scale where the Nernst coefficient approaches zero. While it
should be recognized that this criterion depends on experimental
resolution, it provides a reasonable estimate of the melting field
$B_m$ of the vortex solid.

Those two field scales, $B_m$ and $B_{SIT}$, are reported on the
phase diagram shown on a log scale, figure~\ref{fig:fig8}, panel b),
together with the GCF line obtained from the position of the maximum
in the field dependence of the Nernst data, measured above $T_c$.

This diagrams shows that in the low field limit, the temperature
range for the existence of the vortex liquid is very narrow, and
explains why the temperature dependence of the Nernst coefficient
shows a sharp peak centered at $T_c$, figure ~\ref{fig:fig2}, panel
d). This peak is the consequence of the diverging correlation length
for Cooper pair fluctuations and is not due to the vortex fluid
motion. Just below $T_c$, the Nernst coefficient decreases as the
system enters the vortex solid.

At finite magnetic field, see curve at $B=0.15 T$, figure
~\ref{fig:fig2} panel d), the temperature dependence of the Nernst
coefficient shows a peak that becomes broader than in the zero field
limit as a consequence of the intervening vortex fluid.

\section{Conclusion}

Superconducting fluctuations are at the center of important
contemporary issues in strongly correlated electronic systems. In
cuprates, the identification of the nature of superconducting
fluctuations in the underdoped - high temperature part of the phase
diagram may help elucidating the origin of the pseudo-gap observed
in the electronic spectrum. If so, this will undoubtedly bring us
closer to the solution of the high-T$_{c}$ problem. In amorphous
superconducting thin films, the proper characterization of the
superconducting fluctuations on the insulating side of the quantum
superconductor-insulator transition would shed light on the nature
of this transition and the characteristics of the Bosonic insulator.

This context explains the large attention devoted to the Nernst
effect. While it has been known for a long time to be highly
sensitive to the vortices of the vortex fluid, only recently, did we
discover that it is also highly sensitive to Cooper pair
fluctuations. Theoretically, while the vortex-induced Nernst signal
is exceedingly difficult to analyze as it depends on microscopic
details such as the vortex pinning, the Nernst signal arising from
Cooper-pair fluctuations is simple to analyze as it  only depends on
the size of the superconducting fluctuations. This leads to a simple
relationship between the Nernst coefficient and the superconducting
correlation length, as expressed by USH formula close to $T_c$, and
gives rise to a GCF in the field dependence of the Nernst signal.
Our description of the evolution of the Nernst coefficient across
the superconducting phase diagram of those superconducting films
shows that the examination of unconventional superconducting
fluctuations should be done by considering the deviations with
respect to the Nernst signal generated by Cooper pair fluctuations,
which are expected to exist in any superconductor.

\ack{We thank C. A. Marrache-Kikuchi,  L. Dumoulin and Z. Ovadyahu
who provided us with amorphous superconducting thin films and C.
Capan for the $La_{1.94}Sr_{0.06}CuO_4$ data. The financial support
of the Agence National de la Recherche (ANR-08-BLANC-0121-02) is
acknowledged.}

%
%
%
%
%

\section*{References}
\bibliography{Supra_Isolant}
\end{document}